\documentclass{article}
\usepackage{spconf,amsmath,amssymb,graphicx,booktabs,multirow,xspace}
\usepackage[hidelinks]{hyperref}

\title{TRAINING SPEECH DEEPFAKE DETECTORS AGAINST WORST-CASE FILTERS:\\GAINS, COSTS AND LIMITS}
\name{Ali Alavi}
\address{Department of Computer Science and Engineering, The Ohio State University, Columbus, OH, USA\\ alavibajestan.1@osu.edu}

\begin{document}
\ninept
\maketitle

\begin{abstract}
Speech deepfake detectors make few errors on standard benchmarks, yet a simple attack defeats them: the attacker passes fake speech through one short filter, tuned against the detector, and many fakes then pass as real.
Training with random filters does not stop this attack, because the attacker uses the worst filter, not a random one.
We therefore train the detector against the worst case: during training, a small set of filters is updated to fool the current detector, at about the cost of random filters.
In a preregistered study, this made the detector more robust to filter attacks on real-world recordings than random filters did.
It also lowered the in-domain attack error of the standard detector, and beat one-step adversarial training with additive noise on real-world recordings.
The price is a higher error on unfiltered speech, and a validation study shows that choosing the checkpoint on held-out spoofing systems made the standard detector about as robust.
\end{abstract}


\section{Introduction}
\label{sec:intro}

A speech deepfake is synthetic or converted speech that imitates a real person.
A deepfake detector, also called a spoofing countermeasure, decides whether a recording is bona fide (real) or spoofed~\cite{wang2020asvspoof2019}.
Detectors built on self-supervised speech models reach equal error rates (EERs) below 1\% on standard benchmarks~\cite{tak2022ssl}.
The EER is the error rate at the threshold where false accepts equal false rejects.

These low error rates come with two weaknesses.
First, errors grow on channels not seen in training, such as telephone codecs~\cite{liu2023asvspoof2021} or the unknown processing of audio found online~\cite{muller2022itw}.
Second, an attacker can raise the error on purpose.
Malafide~\cite{panariello2023malafide} learns one short linear filter against a given detector and applies it to the fake speech; the filter only changes the spectral shape of the speech.
The ASVspoof 5 challenge includes such attacks~\cite{wang2024asvspoof5}.

The second weakness is serious.
In our measurements, a Malafide filter raised the EER of a standard XLSR-AASIST detector~\cite{tak2022ssl} from 0.5\% to 14.0\% on the ASVspoof 2019 logical access (LA) evaluation data.
On real-world recordings from In-the-Wild~\cite{muller2022itw}, it raised the EER from 11.0\% to 37.2\%: more than one decision in three was wrong.

The usual defence against channel effects is random augmentation with filters, codecs or noise~\cite{tak2022rawboost,zhang2021channel,ali2024laundering}.
It does not stop filter attacks.
A detector trained with such transforms, including low-pass filtering, still had two Malafide attacks among its five weakest ASVspoof 5 conditions~\cite{ali2024laundering}.
In our study, adding random filters and a consistency loss to the standard recipe left the In-the-Wild attack EER almost unchanged (36.7\% against 37.2\%).
Random filters cover the average case, while the attacker picks the worst filter for the given detector.

We therefore train the detector against the worst case.
During training, each utterance is also passed through filters that are themselves trained to fool the current detector.
Training keeps 8 such filters and updates them with the detector's own backward pass, so it costs about the same as training with random filters.
We compare with random filters of the same length, energy, number of filtered views and cost, in two studies whose hypotheses, data and tests were fixed in time-stamped files before any of their test results existed (preregistration).

\textbf{Relation to prior work.}
Detectors have been made more robust to channels with channel augmentation and domain-adversarial training~\cite{zhang2021channel}, contrastive learning~\cite{wu2024clad} and flat-minimum training~\cite{huang2025sharpness}.
Against attacks~\cite{liu2019adversarial}, the usual defence is adversarial training with additive noise~\cite{wu2020defense,kawa2023defense}.
None of these trains against \emph{learned filters}, the transform Malafide uses against detectors (Malacopula~\cite{todisco2024malacopula} learns non-linear filters against speaker verification).
Our training is universal adversarial training~\cite{shafahi2020universal} with class-wise perturbations~\cite{benz2021classwise}, moved from additive noise to filters; training on worst-case augmentations is also used in domain generalisation~\cite{volpi2018generalizing} and speaker verification~\cite{zhou2024advaug}.
What is new is the bounded filter family, a filter bank whose choice of filter carries no information about the label, and a controlled test against Malafide on real-world data.

\textbf{Contributions.}
(i) Training against a bank of worst-case finite impulse response (FIR) filters, at about the cost of random filters (Sec.~\ref{sec:method}).
(ii) A preregistered result: with a consistency loss, worst-case filters beat random filters of the same energy under attack on real-world recordings, over 10 training seeds (Sec.~\ref{sec:study2}).
(iii) Exploratory comparisons with the standard detector and with one-step adversarial training (Sec.~\ref{sec:baseline}).
(iv) The costs and limits: a higher clean EER, a failed first hypothesis, and a validation study in which better checkpoint choice made the standard detector about as robust (Sec.~\ref{sec:pilot}, \ref{sec:conclusion}).

\section{Worst-case filter training}
\label{sec:method}

\subsection{The filter family and the attack}
Let $f_\theta$ be a detector that maps a waveform $x$ to two scores (spoof, bona fide), and let $\ell$ be its cross-entropy loss.
We use the family of FIR filters
\begin{equation}
  \mathcal{H}_\varepsilon = \{\, h = \delta + v \;:\; v \in \mathbb{R}^{L},\ v_c = 0,\ \|v\|_2^2 \le \varepsilon \,\},
  \label{eq:family}
\end{equation}
where $\delta$ is a unit impulse at the centre tap $c$.
The unit centre tap passes the signal through unchanged, and $v$ adds a spectral change whose energy is at most the energy budget $\varepsilon$.
Malafide~\cite{panariello2023malafide} uses the same form without the energy bound.
For a fixed detector, it finds one filter that makes spoofed utterances score as bona fide: $\min_{h} \mathbb{E}_{x}\, \ell\big(f_\theta(h \ast x), y_{\text{bona fide}}\big)$ over spoofed $x$.
Random augmentation, for example RawBoost~\cite{tak2022rawboost}, instead draws $h$ at random from a similar family.

\subsection{Training against a bank of worst-case filters}
We keep a bank of $K$ filters $h_1, \dots, h_K \in \mathcal{H}_\varepsilon$.
Each filter has a target class $c_k$: half of the bank targets spoofed speech and half targets bona fide speech.
The detector is trained on
\begin{equation}
  \min_\theta \; \mathbb{E}_{(x,y)} \Big[ \ell\big(f_\theta(x), y\big) + \mathbb{E}_{k}\, \ell\big(f_\theta(h_k \ast x), y\big) \Big] + \lambda\, \mathcal{C},
  \label{eq:minmax}
\end{equation}
where $k$ is drawn at random for each example.
Each filter tries to maximise the training loss of the filtered view on examples of its target class.
The spoof-targeted filters therefore learn what Malafide exploits: they push spoofed speech towards the bona fide side.
The bona fide-targeted filters push real speech towards the spoof side.
Every filter is applied to examples of both classes, but only examples of its target class update it.
So the choice of filter carries no information about the label, and the detector cannot use it as a cue for the label.

After each backward pass, every filter used in the batch takes one ascent step.
Let $g_k = \nabla_{v_k} \mathcal{L}$ be the gradient of the filtered-view loss.
The step follows a moving average of the normalised gradients: $m_k \leftarrow \beta m_k + (1-\beta)\, g_k/\|g_k\|_2$ and $v_k \leftarrow v_k + \eta\, m_k/\|m_k\|_2$.
The filter is then projected back onto $\mathcal{H}_\varepsilon$: the centre tap is reset and the energy is clipped to the energy budget.
The same backward pass gives both updates, as in ``free'' adversarial training~\cite{shafahi2019free}.
The only extra cost over random filters is the update of $K \times L$ filter taps per step.
The moving average is needed because a filter sees only a few examples per step.
Without it, the filters stayed at energy $0.06$--$0.19$ at every budget in preliminary runs.
To keep the bank diverse, each filter is re-initialised with probability $p$ per step.

\textbf{Consistency term.}
The \textsc{+C} variants add, with weight $\lambda$, the term
\[ \mathcal{C} = \mathbb{E}\,[1-\cos(e(x), e(h \ast x))], \]
where $e(\cdot)$ is the $160$-dimensional embedding before the output layer.  
The filters also ascend this term, so they also look for spectral changes that move the embedding.

\textbf{Control.}
The matched control (\textsc{Rand}) replaces the bank by a fresh random filter for each example, with Gaussian taps scaled to exactly $\|v\|_2^2 = \varepsilon$.
It has the same filter length, energy budget, number of filtered views and about the same training cost as the worst-case bank (\textsc{Adv}).
The two differ mainly in how the filters are chosen.
They also differ in how long a filter is kept: the 8 filters of the bank are kept for about $1/p$ steps, while \textsc{Rand} draws a fresh filter for each example.

\section{Experimental setup}
\label{sec:setup}

\textbf{Detector and baseline.}
All systems are XLSR-AASIST~\cite{tak2022ssl}: an XLS-R 300M front-end~\cite{babu2022xlsr}, fully fine-tuned, with an AASIST back-end~\cite{jung2022aasist}.
The baseline (RB) is the reference recipe trained on ASVspoof 2019 LA~\cite{wang2020asvspoof2019}: the first $64{,}600$ samples of each utterance (tiled if shorter), RawBoost (convolutive plus impulsive noise)~\cite{tak2022rawboost}, weighted cross-entropy $[0.1, 0.9]$, and Adam with weight decay $10^{-4}$ and batch size 14.  
For compute reasons we train 10 epochs at learning rate $10^{-5}$ instead of 100 epochs at $10^{-6}$, a choice made on validation data before any \textsc{Adv} run finished.  
Our RB reaches $1.58\%$ EER on ASVspoof 2021 LA, against $0.82\%$ for the full recipe~\cite{tak2022ssl}.
The filter methods add one filtered view per utterance; an epoch takes 4.4 minutes for RB, 6.7 for \textsc{Rand+C} and 6.9 for \textsc{Adv+C} on one A100 GPU.
The checkpoint is the epoch with the lowest mean EER over a fixed subset of the ASVspoof 2019 LA development set (2,548 bona fide, 5,000 spoofed), clean and coded with MP3, AAC and Vorbis.  
No codec is used in training, and the validation and test codecs differ.

\textbf{Filter banks.}
$L = 257$ taps, $K = 8$ filters, ascent step $\eta = 0.05$, momentum $\beta = 0.9$, reset probability $p = 0.001$, initial energy $0.01$, and $\lambda = 1$ for \textsc{+C}.  
The energy budget $\varepsilon \in \{0.25, 1, 4\}$ was selected separately for \textsc{Rand} and \textsc{Adv} on the mean of the validation-codec EER and the EER under a Malafide attack built from development-set spoofs.

\textbf{Attacks.}
The attacker knows the trained detector and its weights (white-box access).
For each detector, it learns one FIR filter that is applied to every spoofed utterance (a universal filter), with the official Malafide initialisation, update rule and optimiser~\cite{panariello2023malafide}; we call the attack adaptive because it is trained against the detector under test.
There is no perceptual or energy limit other than Adam's weight decay, and the attack runs for 1,500 steps with batch size 16.  
We use filter lengths 257 and 1025 (the length used in ASVspoof 5~\cite{wang2024asvspoof5}) and report their mean.  
On ASVspoof 2019 LA, the filter is trained on 4,000 evaluation spoofs and scored on 10,000 other spoofs plus all 7,355 bona fide utterances; on In-the-Wild, it is trained on 4,000 spoofs and scored on the other 7,816 spoofs and all 19,963 bona fide utterances.
Transfer attacks apply the filter trained against RB of the same seed, unchanged.

\textbf{Test sets.}
We also test on the ASVspoof 2019 LA evaluation set (in domain), the ASVspoof 2021 LA evaluation subset~\cite{liu2023asvspoof2021} (real telephony and VoIP channels), unseen codecs (Opus narrowband 8\,kbit/s, G.726 16\,kbit/s and EnCodec 6\,kbit/s~\cite{defossez2023encodec}) on 14,710 ASVspoof 2019 LA evaluation utterances, and In-the-Wild~\cite{muller2022itw}.  

\textbf{Protocol.}
Both studies were preregistered with git tags before any of their test scores existed (\texttt{prereg-v1}, commit f53dcd1; \texttt{prereg-v2}, commit e23b77d).
The code locks the test sets and logs every access (one check script was logged by hand).
A first scoring pass stored 16-bit scores, which caused ties; we re-scored all models with 32 bits (Study 1 primary difference $+0.85$ before, $+0.91$ after), the only deviation from the Study 2 preregistration.  
Unless stated otherwise, results are means over seeds 0--4, and differences are paired by seed with 95\% bias-corrected (BCa) bootstrap intervals, which are approximate and tend to be too narrow with 5 seeds.  

\textbf{Validation study.}
After both studies, we ran a validation-only study (3 seeds) on the ASVspoof 5 development set~\cite{wang2024asvspoof5} (first of three archives), whose spoofing systems do not occur in ASVspoof 2019 LA.
Its speakers form two disjoint groups: one with systems A09--A12 for checkpoint selection, and one with systems A13--A16 (4,000 bona fide, 8,000 spoofed; attack trained on 3,000 other spoofs) as the held-out validation set we report.
The held-out selection key is the mean of the usual validation EER, the clean EER on the selection split, and its EER under a short (200-step) Malafide attack.  

\section{Results}
\label{sec:results}

\begin{table}[t]
\centering
\footnotesize
\caption{Study 2 (preregistered): paired difference \textsc{Adv+C} $-$ \textsc{Rand+C} at the same budget (mean over $\varepsilon \in \{0.25, 1\}$), 95\% BCa interval, and seeds with a negative difference. Top: adaptive Malafide on In-the-Wild, never attacked before (seeds 0--9); indented rows are views of the primary result. Bottom: new seeds 5--9; $^\ast$: not a preregistered outcome.}
\label{tab:confirm}
\setlength{\tabcolsep}{3pt}
\begin{tabular}{lcrcc}
\toprule
Metric & Seeds & $\Delta$ & 95\% CI & neg. \\
\midrule
ITW Malafide (primary) & 0--9 & $-$8.02 & [$-$10.57, $-$4.78] & 9/10 \\
\quad new seeds only & 5--9 & $-$7.06 & [$-$8.52, $-$5.60] & 5/5 \\
\quad $\varepsilon = 0.25$ only & 0--9 & $-$8.34 & [$-$12.93, $-$5.39] & 10/10 \\
\quad $\varepsilon = 1$ only & 0--9 & $-$7.69 & [$-$11.55, $-$3.92] & 9/10 \\
\quad increase over no attack & 0--9 & $-$8.97 & [$-$11.24, $-$5.13] & 9/10 \\
\midrule
19LA Malafide & 5--9 & $-$1.08 & [$-$2.62, $+$0.02] & 4/5 \\
Codecs & 5--9 & $+$0.79 & [$+$0.52, $+$1.14] & 0/5 \\
ITW (no attack) & 5--9 & $+$2.14 & [$+$0.84, $+$3.62] & 1/5 \\
21LA$^\ast$ & 5--9 & $+$0.35 & [$+$0.09, $+$0.59] & 1/5 \\
19LA (no attack)$^\ast$ & 5--9 & $+$0.94 & [$+$0.46, $+$1.26] & 0/5 \\
\bottomrule
\end{tabular}
\end{table}

\subsection{Worst-case versus random filters}
\label{sec:study2}
\textbf{Study 1: not supported.}
Both preregistered studies compare \textsc{Adv+C} with \textsc{Rand+C}. Study 1 compared the two systems that did best on validation data: \textsc{Adv+C} with $\varepsilon = 4$ and \textsc{Rand+C} with $\varepsilon = 1$ ($\dagger$ in Table~\ref{tab:all}).
Its main measure $R$ is the mean EER over the attack, ASVspoof 2021 LA, the codecs and In-the-Wild.
Worst-case filters did not win: $R$ was $6.77\%$ against $5.86\%$ ($+0.91$, CI $[-0.71, +1.51]$), and the attack EERs were almost equal ($6.92\%$ against $6.73\%$).
The worst-case system also failed the preregistered test that it is no worse on In-the-Wild ($+1.85$ points).
The test results suggest why: on every test set, \textsc{Adv+C} did better with the smaller budgets $\varepsilon \in \{0.25, 1\}$ than with $\varepsilon = 4$.
One possible reason is that the validation attack used the spoofing systems seen in training (A01--A06), while the test attack uses unseen ones (A07--A19).
Since this pattern came from the test results, we tested it again in Study 2.

\textbf{Study 2: supported.}
We preregistered Study 2 after Study 1 failed and before any of its data existed.
It uses the same budget for both, $\varepsilon \in \{0.25, 1\}$, five new models (seeds 5--9) and a test that no model had faced before: the adaptive Malafide attack on In-the-Wild.  
Worst-case filters won (Table~\ref{tab:confirm}): under attack, the In-the-Wild EER was $28.7\%$ with worst-case filters and $36.7\%$ with random filters, 8.02 points lower (CI $[-10.57, -4.78]$), and it was lower for 9 of 10 seeds.  
It also holds on the new seeds alone and at each budget (Table~\ref{tab:confirm}).
It does not come from a difference in clean error: the attack added 15.1 points of EER for \textsc{Adv+C} but 24.1 points for \textsc{Rand+C}.  
On the in-domain attack, the two differed little on the new seeds ($-1.08$, CI $[-2.62, +0.02]$).  
The gain has a price: on the new seeds, worst-case filters raised the EER on unseen codecs by 0.79 points and on clean In-the-Wild speech by 2.14 points.
Without the consistency loss ($\lambda = 0$), worst-case and random filters did not differ reliably on the in-domain attack (three matched pairs in Table~\ref{tab:all}: $+0.53$ to $-2.42$ points, every interval includes zero), and we did not test $\lambda = 0$ on In-the-Wild, so our claims cover only the consistency-loss variants.

\begin{table}[t]
\centering
\footnotesize
\caption{Comparison with the baseline (RB): test EER (\%), mean over seeds 0--4. Attacked: adaptive Malafide on 19LA and ITW. Codecs: three unseen codecs. \textsc{Adv+C}$^\dagger$: preregistered Study 1 system ($\varepsilon{=}4$). \textsc{Rand+C}, \textsc{Adv+C}: mean over $\varepsilon{\in}\{0.25,1\}$, chosen after seeing the 19LA test results. $^\ddagger$: one-step adversarial training with additive noise (30\,dB, not tuned), run last. $\Delta$: paired difference to RB (seeds with $\Delta{<}0$). Only $\Delta$ \textsc{Adv+C}$^\dagger$ on 19LA was preregistered.}
\label{tab:baseline}
\setlength{\tabcolsep}{2.2pt}
\begin{tabular}{lcc|cccc}
\toprule
 & \multicolumn{2}{c|}{Attacked} & \multicolumn{4}{c}{Clean} \\
System & 19LA & ITW & 19LA & 21LA & Codecs & ITW \\
\midrule
RB (baseline) & 14.0 & 37.2 & 0.5 & 1.6 & 1.1 & 11.0 \\
\textsc{Adv+C}$^\dagger$ & 6.9 & -- & 2.6 & 2.7 & 2.6 & 14.8 \\
\textsc{Rand+C} & 8.7 & 36.7 & 1.6 & 2.2 & 1.5 & 13.2 \\
\textsc{Adv+C} & 5.1 & 27.8 & 1.5 & 2.2 & 2.0 & 13.1 \\
Additive adv.$^\ddagger$ & 6.4 & 44.9 & 1.4 & 2.2 & 1.6 & 11.5 \\
\midrule
$\Delta$ \textsc{Adv+C}$^\dagger$ & $-$7.1\,{\scriptsize(5)} & -- & $+$2.1\,{\scriptsize(0)} & $+$1.1\,{\scriptsize(0)} & $+$1.5\,{\scriptsize(0)} & $+$3.8\,{\scriptsize(0)} \\
$\Delta$ \textsc{Rand+C} & $-$5.3\,{\scriptsize(5)} & $-$0.5\,{\scriptsize(3)} & $+$1.1\,{\scriptsize(0)} & $+$0.6\,{\scriptsize(0)} & $+$0.4\,{\scriptsize(1)} & $+$2.2\,{\scriptsize(0)} \\
$\Delta$ \textsc{Adv+C} & $-$8.9\,{\scriptsize(5)} & $-$9.5\,{\scriptsize(4)} & $+$1.0\,{\scriptsize(0)} & $+$0.6\,{\scriptsize(0)} & $+$0.9\,{\scriptsize(0)} & $+$2.1\,{\scriptsize(1)} \\
$\Delta$ Additive adv.$^\ddagger$ & $-$7.6\,{\scriptsize(4)} & $+$7.7\,{\scriptsize(1)} & $+$0.9\,{\scriptsize(0)} & $+$0.6\,{\scriptsize(0)} & $+$0.5\,{\scriptsize(0)} & $+$0.5\,{\scriptsize(2)} \\
\bottomrule
\end{tabular}
\end{table}

\subsection{Comparison with the baseline}
\label{sec:baseline}
Only one comparison with the standard detector RB was preregistered: the Study 1 system (\textsc{Adv+C}$^\dagger$) on the in-domain attack (Table~\ref{tab:baseline}).
It halved the attack EER, from 14.0\% to 6.9\% ($-7.1$ points, CI $[-10.2, -4.5]$, lower for all 5 seeds).
Its clean EER, however, rose by 1.1 to 3.8 points, so its mean EER $R$ did not improve ($-0.15$, CI $[-1.67, +1.18]$).  

The other rows are exploratory.
They use the Study 2 budgets, which we chose after seeing the in-domain test results, so their in-domain numbers are optimistic: on the new seeds 5--9, the in-domain attack EER of \textsc{Adv+C} is 7.4\%, not 5.1\%.  
On In-the-Wild under attack, worst-case filters reached 27.8\% against 37.2\% for RB on seeds 0--4 (28.7\% over all ten seeds), while random filters gave no gain (36.7\%).  
This drop of 9.5 points appears in 4 of 5 seeds, but five seeds are too few to call it reliable (t-test $p = 0.13$).
The price is clean accuracy: all three filter-trained systems in Table~\ref{tab:baseline} had a higher clean EER than RB on every test set.

\textbf{Adversarial training with additive noise.}
This is the usual defence against attacks on detectors~\cite{wu2020defense,kawa2023defense}.
We ran one untuned version, simpler than the multi-attack training of~\cite{kawa2023defense}: one-step adversarial noise at 30\,dB, with the same consistency loss, recipe and attacks ($^\ddagger$ in Table~\ref{tab:baseline}; exploratory, run last).
In domain it was about as robust as worst-case filters (6.4\% against 5.1\%).  
On In-the-Wild under attack, however, it reached 44.9\%: worse than RB (37.2\%; $+7.7$ points, CI $[+1.0, +14.3]$) and 17.1 points worse than worst-case filters (CI $[+11.4, +23.1]$, worse for all 5 seeds).  
Its clean EERs were equal to or lower than those of worst-case filters, most clearly on In-the-Wild (11.5\% against 13.1\%, all 5 seeds).  
So training against additive noise did not protect against filters on real-world recordings, while training against worst-case filters partly did.

For all twelve filter-trained configurations in Table~\ref{tab:all}, attacks transferred from RB were weaker than the direct attack, a check against gradient masking (false robustness from broken gradients).

\begin{table*}[t]
\centering
\footnotesize
\caption{Test EER (\%) of every configuration: mean over 5 seeds, seed standard deviation as subscript. 19LA: ASVspoof 2019 LA eval (in domain). 21LA: ASVspoof 2021 LA eval subset (telephony and VoIP channels). Codecs: mean over unseen Opus-NB 8k, G.726 16k and EnCodec 6k. ITW: In-the-Wild. Mal.: adaptive white-box Malafide (mean over filter lengths 257 and 1025). Transf.: Malafide filters trained against RB of the same seed, applied unchanged. $R$: preregistered primary metric (mean of 21LA, Codecs, ITW, Mal.). $\dagger$: the preregistered proposed/control pair of Study 1, chosen on validation data.}
\label{tab:all}
\setlength{\tabcolsep}{3.5pt}
\begin{tabular}{llcccccccc}
\toprule
Method & $\lambda$ & $\varepsilon$ & 19LA & 21LA & Codecs & ITW & Mal. & Transf. & $R$ \\
\midrule
RB & -- & -- & 0.49$_{\pm 0.1}$ & 1.58$_{\pm 0.1}$ & 1.12$_{\pm 0.1}$ & 11.00$_{\pm 1.4}$ & 13.99$_{\pm 4.1}$ & -- & 6.92$_{\pm 1.2}$ \\
\midrule
Rand & 0 & 0.25 & 0.88$_{\pm 0.3}$ & 1.63$_{\pm 0.2}$ & 1.13$_{\pm 0.2}$ & 10.03$_{\pm 3.9}$ & 7.84$_{\pm 3.7}$ & 7.00$_{\pm 4.0}$ & 5.16$_{\pm 0.8}$ \\
Rand & 0 & 1 & 1.95$_{\pm 0.9}$ & 2.70$_{\pm 0.5}$ & 1.94$_{\pm 0.4}$ & 15.23$_{\pm 2.2}$ & 9.86$_{\pm 2.4}$ & 5.27$_{\pm 2.6}$ & 7.43$_{\pm 1.3}$ \\
Rand & 0 & 4 & 1.21$_{\pm 0.3}$ & 1.97$_{\pm 0.4}$ & 1.68$_{\pm 0.5}$ & 14.76$_{\pm 2.1}$ & 8.28$_{\pm 3.6}$ & 4.97$_{\pm 2.7}$ & 6.67$_{\pm 1.3}$ \\
\midrule
Adv & 0 & 0.25 & 1.38$_{\pm 0.4}$ & 2.39$_{\pm 0.6}$ & 2.20$_{\pm 0.8}$ & 11.31$_{\pm 1.5}$ & 8.37$_{\pm 3.2}$ & 5.39$_{\pm 3.0}$ & 6.07$_{\pm 0.9}$ \\
Adv & 0 & 1 & 2.85$_{\pm 1.7}$ & 2.98$_{\pm 0.7}$ & 3.00$_{\pm 0.5}$ & 13.47$_{\pm 1.4}$ & 7.44$_{\pm 4.9}$ & 6.87$_{\pm 3.1}$ & 6.73$_{\pm 1.5}$ \\
Adv & 0 & 4 & 1.93$_{\pm 0.9}$ & 2.64$_{\pm 0.4}$ & 2.43$_{\pm 0.5}$ & 14.30$_{\pm 2.1}$ & 6.33$_{\pm 4.1}$ & 4.76$_{\pm 3.1}$ & 6.43$_{\pm 1.0}$ \\
\midrule
Rand+C & 1 & 0.25 & 1.84$_{\pm 1.2}$ & 2.08$_{\pm 0.3}$ & 1.48$_{\pm 0.4}$ & 13.38$_{\pm 1.1}$ & 10.65$_{\pm 4.0}$ & 8.72$_{\pm 6.0}$ & 6.90$_{\pm 1.1}$ \\
Rand+C$^\dagger$ & 1 & 1 & 1.36$_{\pm 0.5}$ & 2.22$_{\pm 0.4}$ & 1.51$_{\pm 0.4}$ & 12.99$_{\pm 1.7}$ & 6.73$_{\pm 1.4}$ & 5.17$_{\pm 2.8}$ & 5.86$_{\pm 0.6}$ \\
Rand+C & 1 & 4 & 1.95$_{\pm 0.8}$ & 2.36$_{\pm 0.3}$ & 1.73$_{\pm 0.4}$ & 15.33$_{\pm 0.6}$ & 7.52$_{\pm 3.6}$ & 5.71$_{\pm 3.0}$ & 6.73$_{\pm 1.0}$ \\
\midrule
Adv+C & 1 & 0.25 & 1.50$_{\pm 0.6}$ & 2.27$_{\pm 0.5}$ & 2.00$_{\pm 0.5}$ & 13.43$_{\pm 1.3}$ & 4.92$_{\pm 1.8}$ & 3.87$_{\pm 1.9}$ & 5.65$_{\pm 0.8}$ \\
Adv+C & 1 & 1 & 1.41$_{\pm 0.4}$ & 2.19$_{\pm 0.3}$ & 2.02$_{\pm 0.7}$ & 12.82$_{\pm 1.7}$ & 5.30$_{\pm 2.3}$ & 3.40$_{\pm 1.6}$ & 5.58$_{\pm 1.1}$ \\
Adv+C$^\dagger$ & 1 & 4 & 2.64$_{\pm 1.1}$ & 2.70$_{\pm 0.5}$ & 2.62$_{\pm 0.6}$ & 14.84$_{\pm 2.9}$ & 6.92$_{\pm 3.0}$ & 4.30$_{\pm 2.7}$ & 6.77$_{\pm 1.3}$ \\
\bottomrule
\end{tabular}
\end{table*}

\begin{table}[t]
\centering
\footnotesize
\caption{Validation study (exploratory, seeds 0--2). Held-out: ASVspoof 5 dev, systems A13--A16. Mal.: adaptive Malafide. Ep.: selected epoch. Mcop.: learned non-linear (Hammerstein) filter, as in Malacopula. Additive adv.: as in Table~\ref{tab:baseline}. Top: usual checkpoint rule. Bottom: held-out checkpoint rule (Sec.~\ref{sec:setup}). $^\ast$From the 20-epoch run.}
\label{tab:pilot}
\setlength{\tabcolsep}{3pt}
\begin{tabular}{lccccc}
\toprule
 & & \multicolumn{3}{c}{Held-out} & 19LA val \\
System & Ep. & Mal. & Mcop. & Clean & Mal. \\
\midrule
RB (baseline) & 6.7 & 47.4 & 57.2 & 10.5 & 18.2 \\
\textsc{Rand+C}, $\varepsilon{=}1$ & 5.3 & 33.7 & 35.0 & 13.3 & 3.5 \\
\textsc{Adv+C}, $\varepsilon{=}1$ & 7.0 & 31.7 & 29.9 & 13.0 & 9.5 \\
\textsc{Adv+C}, $\varepsilon{=}4$ & 4.3 & 22.9 & 22.0 & 11.0 & 3.2 \\
Additive adv. & 6.0 & 33.7 & 50.0 & 14.8 & 7.7 \\
\midrule
RB$^\ast$ & 1.0 & 18.3 & 20.8 & 5.1 & 6.0 \\
\textsc{Rand+C}, $\varepsilon{=}1$ & 2.3 & 25.2 & 24.7 & 9.3 & 5.8 \\
\textsc{Adv+C}, $\varepsilon{=}1$ & 1.7 & 19.5 & 21.1 & 10.5 & 3.7 \\
Additive adv. & 1.3 & 22.5 & 22.6 & 7.4 & 6.9 \\
\bottomrule
\end{tabular}
\end{table}

\subsection{Validation study on unseen spoofing systems}
\label{sec:pilot}
With the usual rule for choosing the checkpoint (Table~\ref{tab:pilot}, top), filter training helped against the held-out attack: the EER fell from 47.4\% for RB to 33.7\% with random filters and 31.7\% with worst-case filters at $\varepsilon = 1$.
Worst-case filters were only slightly better than random ones (about 2 points, lower in 2 of 3 seeds).

How the checkpoint is chosen changed the results more than any training change we tried (Table~\ref{tab:pilot}, bottom).
Over all 48 runs of the study, most of them variants not shown in Table~\ref{tab:pilot}, choosing it on held-out spoofing systems picked much earlier epochs (1.6 instead of 7.1 on average), lowered the held-out attack EER by 9.9 points and the clean EER by 5.4 points, at a small cost on the in-domain validation data ($+0.5$ points clean, $+0.9$ points under attack).
With this rule, RB reached 18.3\% under attack, about the same as worst-case filters (19.5\%), with half their clean EER (5.1\% against 10.5\%).
This RB comes from a 20-epoch run, because the 10-epoch runs stored no held-out checkpoint; its learning rate in the first three epochs is within 6\% of the standard recipe.
Worst-case filters still beat random filters, in all three seeds ($-5.7$ points).

\textbf{Controls.}
A fixed bank of 8 random filters that is never updated matched the worst-case bank under Malafide at the usual checkpoint ($+0.3$ points), so there the small edge over fresh random filters comes from keeping filters longer; but it was 10.3 points worse under the Malacopula-style attack, and 7.6 points worse under Malafide with the held-out rule.
A second RawBoost view without filters was 23.0 points worse than random filters.
On validation, additive adversarial training was about 20 points worse than worst-case filters under the Malacopula-style attack at the usual checkpoint, but close to them with the held-out rule (Table~\ref{tab:pilot}).  
Six other changes failed a screening rule fixed before the runs.
These are three-seed validation results, and one corpus was used to choose and to report, so part of the held-out rule's gain may be specific to ASVspoof 5.

\subsection{Checks}
\label{sec:checks}
A detector that rejects all filtered audio would also do well against Malafide, which filters only fake speech.
With the attack filter applied to real speech too, the two systems had similar EERs ($24.7\%$ for \textsc{Adv+C}, $25.5\%$ for \textsc{Rand+C}, 40 models), so \textsc{Adv+C} does not work this way.  

The attack is not fully converged at 1,500 steps: with 4,500 steps (seeds 5--6), the EER rises from $32.3\%$ to $36.1\%$ for \textsc{Adv+C} and from $38.8\%$ to $41.8\%$ for \textsc{Rand+C}, and the gap narrows only from $6.5$ to $5.7$ points.  

\section{Conclusion}
\label{sec:conclusion}
Filter attacks are cheap and strong: one short learned filter makes a standard detector wrong on more than one real-world decision in three.
Random filter augmentation does not stop them, because the attacker uses the worst filter.
Training against a small bank of worst-case filters costs about the same; in a preregistered study, it made the detector more robust to filter attacks on real-world recordings than random filters of the same energy, and in exploratory tests it also beat one-step adversarial training with additive noise there, at some cost in accuracy on unfiltered speech.
Our validation study points to a simpler lever too: choosing the checkpoint on held-out spoofing systems made the standard detector about as robust; testing it on the test sets comes next.

\textbf{Limitations.}
We tested one detector, trained on one corpus with a shortened recipe, and only filter attacks on the test sets, each run once per model (Sec.~\ref{sec:checks}).
We did not use the ASVspoof 5 evaluation conditions built for filter attacks~\cite{wang2024asvspoof5}, and the additive-noise baseline is one untuned setting.
Code, configurations, preregistrations and logs will be released.

\section{Compliance with ethical standards}
This research study was conducted retrospectively using human subject data made available in open access by ASVspoof 2019, 2021 and 5 and by In-the-Wild. Ethical approval was not required as confirmed by the licenses attached with the open access data.

\bibliographystyle{IEEEbib}
\bibliography{refs}

\end{document}